\documentclass[12pt, a4paper]{article}
\usepackage{a4,amsfonts,amssymb,amsmath,amsthm,graphicx}

\newcommand{\BB}{\vert_{\partial M}}
\newcommand{\tr}{{\rm tr}\,}
\newcommand{\Dir}{\widehat D}
\newcommand{\iM}{\int_M d^nx\sqrt{g}}
\newcommand{\idM}{\int_{\partial M} d^{n-1}x\sqrt{h}}

\begin{document}

\title{Spectral functions and their applications}

\author{Valery N.Marachevsky \thanks{email: maraval@mail.ru,
root@VM1485.spb.edu} \\
{\it V. A. Fock Institute of Physics, St. Petersburg
University,}\\
{\it 198504 St. Petersburg, Russia} }



\maketitle

\begin{abstract}
We give an introduction to the heat kernel technique and
$\zeta$-function. Two applications are considered. First we derive
the high temperature asymptotics of the free energy for boson
fields in terms of the heat kernel expansion and $\zeta$-function.
Another application is chiral anomaly for local (MIT bag) boundary
conditions.
\end{abstract}

\maketitle

\section{Introduction}

In this paper we give an introduction to the technique of the
heat-kernel expansion and $\zeta$-function regularization. The
heat kernel became a standard tool in calculations of the vacuum
polarization, the Casimir effect and quantum anomalies. In the
case of non-trivial background fields, and especially in the
presence of boundaries or singularities, the heat kernel technique
seems to be the most adequate one for the analysis of the one-loop
effects (see \cite{Vassilevich:2003xt} for a recent review).

A possible application of heat kernel methods to the evaluation of
high temperature asymptotics of the free energy in the presense of
boundaries or singularities has not received due attention yet. At
high temperatures the expansion of the free energy can be
determined by the heat kernel expansion and $\zeta$-function
\cite{Dowker}.

Chiral anomaly, which was discovered more than 30 years ago
\cite{abj}, still plays an important role in physics. On smooth
manifolds without boundaries many successful approaches to the
anomalies exist \cite{Bertlmann}.  The heat kernel approach to the
anomalies is essentially equivalent to the Fujikawa approach
\cite{Fujikawa:1979ay} and to the calculations based on the
finite-mode regularization \cite{Andrianov:1983fg}, but it can be
more easily extended to complicated geometries. The chiral anomaly
in the case of non-trivial boundary conditions (MIT bag boundary
conditions) has been calculated only recently \cite{lastpaper}.

The paper is organized as follows. In  Sec. $2$ we give an
introduction to the formalism of the heat kernel and heat kernel
expansion. Also we introduce a zeta function and calculate the
one-loop effective action in terms of the zeta function. In Sec.
$3$ we consider two examples. First we derive the high temperature
expansion of the free energy for boson fields in terms of the heat
kernel expansion and $\zeta$-function. Then we discuss a chiral
anomaly in four dimensions for an euclidean version of the MIT bag
boundary conditions \cite{bag}.

\section{Spectral functions}
\subsection{Heat kernel}

Consider a second order elliptic partial differential operator $L$
of Laplace type on an n-dimensional Riemannian manifold. Any
operator of this type can be expanded locally as
\begin{equation}
  L = - (g^{\mu\nu} \partial_\mu \partial_\nu + a^\sigma \partial_\sigma + b ),
  \label{B4}
\end{equation}
where $a$ and $b$ are some matrix valued functions and
$g^{\mu\nu}$ is the inverse metric tensor on the manifold. For a
flat space $g^{\mu\nu}=\delta^{\mu\nu}$.

The heat kernel can be defined as follows:
\begin{equation}
K(t; x; y; L) = \langle x | \exp (-tL) | y \rangle =
\sum_{\lambda} \phi_{\lambda}^{\dagger} (x) \phi_{\lambda} (y)
\exp (-t\lambda)  ,
\end{equation}
where $\phi_{\lambda}$ is an eigenfunction of the operator $L$
with the eigenvalue $\lambda$.

It satisfies the heat equation
\begin{equation}
(\partial_t + L_x) K (t; x ; y ; L) = 0
\end{equation}
with an initial condition
\begin{equation}
K(0; x; y; L) = \delta (x, y) .
\end{equation}

If we consider the fields in a finite volume then it is necessary
to specify boundary conditions. Different choices are possible. In
section $3.1$ we will consider the case of periodic boundary
conditions on imaginary time coordinate, which are specific for
boson fields. In section $3.2$ we will study bag boundary
conditions imposed on fermion fields. If the normal to the
boundary component of the fermion current $\psi^\dag \gamma_n
\psi$ vanishes at the boundary, one can impose bag boundary
conditions, a particular case of mixed boundary conditions. We
assume given two complementary projectors $\Pi_\pm$,
$\Pi_-+\Pi_+=I$ acting on a multi component field (the
eigenfunction of the operator $L$) at each point of the boundary
and define mixed boundary conditions by the relations
\begin{equation}
\Pi_-\psi \BB =0\,,\quad \left( \nabla_n + S\right) \Pi_+ \psi \BB
=0 \,, \label{mixedbc}
\end{equation}
where $S$ is a matrix valued function on the boundary. In other
words, the components $\Pi_-\psi$ satisfy Dirichlet boundary
conditions, and $\Pi_+\psi$ satisfy Robin (modified Neumann) ones.

It is convenient to define
\begin{equation}
\chi =\Pi_+ - \Pi_- \,.\label{defchi}
\end{equation}

Let $\{ e_j \}$, $j=1,\dots,n$ be a local orthonormal frame for
the tangent space to the manifold and let on the boundary $e_n$ be
an inward pointing normal vector.

The extrinsic curvature is defined by the equation
\begin{equation}
L_{ab}=\Gamma_{ab}^n \,,\label{Lab}
\end{equation}
where $\Gamma$ is the Christoffel symbol. For example, on the unit
sphere $S^{n-1}$ which bounds the unit ball in $R^n$ the extrinsic
curvature is $L_{ab}=\delta_{ab}$.

Curved space offers no complications in our approach compared to
the flat case. Let $R_{\mu\nu\rho\sigma}$ be the Riemann tensor,
and let $R_{\mu\nu}={R^\sigma}_{\mu\nu\sigma}$ be the Ricci
tensor. With our sign convention the scalar curvature
$R=R_\mu^\mu$ is $+2$ on the unit sphere $S^2$. In flat space the
Riemann and Ricci tensors are equal to zero.

One can always introduce a connection $\omega_\mu$ and another
matrix valued function $E $ so that $L$ takes the form:
\begin{equation}
  L= - (g^{\mu\nu}\nabla_\mu \nabla_\nu + E)  \label{B5}
\end{equation}
Here $\nabla_\mu $ is a sum of covariant Riemannian derivative
with respect to metric $g_{\mu\nu}$ and connection $\omega_\mu$.
One can, of course, express $E$ and $\omega$ in terms of $a^\mu$,
$b$ and $g_{\mu\nu}$:

\begin{eqnarray}
&&\omega_\mu = \frac{1}{2} g_{\mu\nu} (a^\nu +
  g^{\rho\sigma}\Gamma_{\rho\sigma}^\nu) , \label{B6}
\\
&& E = b - g^{\mu\nu}(\partial_\nu \omega_\mu +
\omega_\mu\omega_\nu
  -\omega_\rho\Gamma_{\mu\nu}^\rho )    \label{B7}
\end{eqnarray}

For the future use we introduce also the field strength for
$\omega$:
\begin{equation}
 \Omega_{\mu\nu}=\partial_\mu\omega_\nu -
\partial_\nu\omega_\mu + [\omega_\mu, \omega_\nu] \,.\label{Omega}
\end{equation}

The connection $\omega_\mu$ will be used to construct covariant
derivatives.  The subscript $;\mu\dots \nu\sigma$ will be used to
denote repeated covariant derivatives with the connection $\omega$
and the Christoffel connection on $M$. The subscript $:a\dots b c$
will denote repeated covariant derivatives containing $\omega$ and
the Christoffel connection on the boundary. Difference between
these two covariant derivatives is measured by the extrinsic
curvature (\ref{Lab}). For example,
$E_{;ab}=E_{:ab}-L_{ab}E_{;n}$.

Let us define an integrated heat kernel for a hermitian operator
$L$ by the equation:
\begin{equation}
K(Q,L,t):={\rm Tr} \left( Q \exp (-tL) \right) = \iM \tr \left(
Q(x) K(t;x;x;L) \right) \,,\label{defhk}
\end{equation}
where $Q(x)$ is an hermitian matrix valued function, ${\rm tr}$
here is over matrix indices. For the boundary conditions we
consider in this paper there exists an asymptotic expansion
\cite{Gilkey:1994} as $t\to 0$:
\begin{equation}
K(Q,L,t) \simeq \sum_{k=0}^\infty a_k (Q,L) t^{(k-n)/2} \,.
\label{hkexp}
\end{equation}

According to the general theory \cite{Gilkey:1994} the
coefficients $a_k(Q,L)$ are locally computable. This means that
each $a_k(Q,L)$ can be represented as a sum of volume and boundary
integrals of local invariants constructed from $Q$, $\Omega$, $E$,
the curvature tensor, and their derivatives. Boundary invariants
may also include $S$, $L_{ab}$ and $\chi$. Total mass dimension of
such invariants should be $k$ for the volume terms and $k-1$ for
the boundary ones.

At the moment several coefficients of the expansion (\ref{hkexp})
are known for the case of mixed boundary conditions
(\ref{mixedbc}) and matrix valued function $Q$ (see
\cite{lastpaper} for details of derivation; the formula
(\ref{Qa4bou}) for $a_4$ was derived in \cite{lastpaper} with
additional restrictions $L_{ab}=0$ and $S=0$) :
\begin{eqnarray}
&&a_0(Q,L)=(4\pi)^{-n/2}\iM \,{\rm tr}\,(Q).\label{Qa0bou} \\
&&a_1(Q,L)={\frac 14}(4\pi)^{-(n-1)/2}\idM
     \,{\rm tr}\, (\chi Q). \label{Qa1bou} \\
&&a_2(Q,L)=\frac 16 (4\pi)^{-n/2}\left\{ \iM
     \,{\rm tr}\,(6QE+QR) \right. \nonumber \\
&&\qquad\qquad \left. +\idM \,{\rm tr} \,
    (2Q L_{aa} +12QS +3 \chi Q_{ ;n})
     \right\} .\label{Qa2bou} \\
&&a_3(Q,L)=\frac 1{384}(4 \pi )^{
       -(n-1)/2}  \idM
       {\rm tr} \big\{ Q( -24 E + 24 \chi E \chi \nonumber\\
&&\qquad\qquad +48 \chi E + 48 E\chi
                     -12 \chi_{ :a} \chi_{:a} + 12 \chi_{:aa}
              -6 \chi_{:a}\chi_{:a}\chi +
              16\chi  R
\nonumber \\ &&\qquad\qquad + 8 \chi R_{anan} +192 S^2 + 96 L_{aa}
S + (3+10\chi )L_{aa}L_{bb} \nonumber \\ &&\qquad\qquad +(6-4\chi
) L_{ab}L_{ab} ) + Q_{;n}(96 S +192 S^2)
              +24 \chi Q_{ ;nn} \big\}.\label{Qa3bou}
\end{eqnarray}
For a scalar function $Q$ and mixed boundary conditions the
coefficients $a_4$ and $a_5$ were already derived \cite{a4a5}.

\subsection{$\zeta$-function}

 Zeta function of an operator $L$ is defined by
\begin{equation}
\zeta_{L}(s) = \sum_\lambda \frac{1}{\lambda^s} \, ,
\end{equation}
where the sum is over all eigenvalues of the operator $L$. The
 zeta function is related to the heat kernel by the transformation
\begin{equation}
\zeta_{L}(s) = \frac{1}{\Gamma(s)} \int_0^{+\infty} dt \, t^{s-1}
K(I, L, t) \, .
\end{equation}
Residues at the poles of the zeta function are related to the
coefficients of the heat kernel expansion:
\begin{equation}
a_k (I, L) = {\rm Res}_{s=(n-k)/2} (\Gamma(s) \zeta_{L}(s) ) \,
.\label{res}
\end{equation}
Here $I$ is a unit matrix with a dimension of the matrix functions
$a^\mu, b$ in (\ref{B4}).From (\ref{res}) it follows that
\begin{equation}
a_n (I, L) = \zeta_{L}(0) \, .
\end{equation}

In  Euclidean four dimensional space the zero temperature one-loop
path integral over the boson fields $\phi = \sum_\lambda C_\lambda
\phi_\lambda$ can be evaluated as follows (up to a normalization
factor):
\begin{equation}
Z = \int d \phi e^{- \int d^4 x \, \phi L\, \phi} \simeq
\prod_\lambda \int \mu d C_\lambda e^{-\lambda C_{\lambda}^2}
\simeq \mu^{\zeta_{L}(0)} {\rm det} L^{-1/2} \, .
\end{equation}
Here we introduced the constant $\mu$ with a dimension of mass in
order to keep a proper dimension of the measure in the functional
integral. $\zeta_L(0)$ can  be thought of as a number of
eigenvalues of the operator $L$. For the operator $L$ in the form
(\ref{B4}) the number of eigenvalues is infinite, so $\zeta_L(0)$
yields a regularized value for this number.

The zero temperature one-loop effective action is defined then by
\begin{eqnarray}
&& W = - \ln Z = -\frac{1}{2} \ln {\rm det} L + \frac{1}{2}
\zeta_{L}(0) \ln \mu^2 = \frac{1}{2} \zeta_{L}^\prime (0) +
\frac{1}{2}
\zeta_L(0) \ln \mu^2 = \nonumber \\
  && \, \quad = \frac{1}{2}
  \frac{\partial}{\partial s} (\mu^{2s}\zeta_{L}(s))|_{s=0}
\end{eqnarray}

The term $\zeta_L(0) \ln \mu^2 = a_4 (I,L) \ln \mu^2$ in the
effective action $W$ determines the one-loop beta function,  this
term describes renormalization of the one-loop logarithmic
divergences appearing in the theory.

\section{Applications}
\subsection{Free energy for boson fields}

A finite temperature field theory is defined in Euclidean space,
since for boson fields one has to impose periodic boundary
conditions on imaginary time coordinate (antiperiodic boundary
conditions for fermion fields respectively). A partition function
is defined by
\begin{equation}
Z(\beta) = {\rm Tr}\, e^{-\beta H} \, ,
\end{equation}
where $H$ is a hamiltonian of the problem and $\beta=\hbar/T$. Let
us choose the lagrangian density $\rho$ in the form
\begin{equation}
\rho= -\frac{\partial^2}{\partial \tau^2} + L \, ,
\end{equation}
where $\tau$ is an imaginary time coordinate and $L$ is a three
dimensional spatial part of the density in the form (\ref{B4}).
The free energy of the system is defined by
\begin{equation}
F (\beta) = - \frac{\hbar}{\beta} \ln Z(\beta) =
 - \frac{\hbar}{\beta} \ln \Bigl( N_\beta \int D\phi
 \exp \bigl( -\int_0^\beta d\tau \int d^3 x \,   \phi \rho \phi \bigr) \Bigr)  \,
 ,
\end{equation}
the integration is over all periodic fields satisfying $\phi(\tau
+ \beta) = \phi(\tau)$ ($N_\beta$ is a normalization coefficient).
As a result the eigenfunctions of $\rho$ have the form
$\exp(i\tau\omega_n)\phi_\lambda$, where $\omega_n = 2\pi n /
\beta$ and $L \phi_\lambda = \lambda \phi_\lambda$. The free
energy is thus equal to \cite{Bordag}
\begin{equation}
F =  \frac{\hbar}{2\beta} \sum_{n=-\infty}^{+\infty} \sum_\lambda
\ln \frac{(\omega_n^2 + \lambda)}{\mu^2} = - \frac{\hbar}{2\beta}
\frac{\partial}{\partial s} (\mu^{2s}\zeta(s) )|_{s=0} \, ,
\end{equation}
where we introduced $\zeta$-function
\begin{equation}
\zeta(s) = \sum_{n=-\infty}^{+\infty} \sum_{\lambda} (\omega_n^2 +
\lambda) ^{-s}
\end{equation}
and the parameter $\mu$ with a mass dimensionality in order to
make the argument of the logarithm dimensionless (also see a
previous section).

Then it is convenient to use the formula
\begin{equation}
\zeta(s) = \frac{1}{\Gamma(s)} \int_0^{+\infty} dt \, t^{s-1}
\sum_{n=-\infty}^{+\infty} \sum_{\lambda} e^{-t (\omega_n^2 +
\lambda)} \, ,
\end{equation}
and separate $n=0$ and other terms in the sum. For $n\neq 0$
terms we substitute the heat kernel expansion for the operator $L$
at small $t$
\begin{equation}
\sum_{\lambda} e^{-\lambda t} = K(I;L;t) \simeq \sum_{k=0}^\infty
a_k (I,L) t^{(k-3)/2}
\end{equation}
and perform $t$ integration, then we arrive at the high
temperature expansion ($\beta \to 0$) for the free energy $F$:
\begin{eqnarray}
&& F/\hbar = - \frac{1}{2\beta} \zeta_L^\prime(0)
-\frac{1}{2\beta} \zeta_L(0) \ln(\mu^2) + (4\pi)^{3/2} \biggl[
-\frac{a_0}{\beta^4} \frac{\pi^2}{90} - \frac{a_{1}}{\beta^3}
\frac{\zeta_R(3)}{4\pi^{3/2}} -\frac{a_2}{\beta^2} \frac{1}{24}
\nonumber \\ && + \frac{a_3}{\beta} \frac{1}{(4\pi)^{3/2}}
\ln\Bigl(\frac{\beta\mu} {2\pi}\Bigr)
-\frac{a_4}{16\pi^2}\biggl(\gamma + \ln \frac{\beta\mu}{2\pi}
\biggr) \nonumber \\ && - \sum_{n\ge 5} \frac{a_n}{\beta^{4-n}}
\frac{(2\pi)^{3/2-n}}{2\sqrt{2}} \Gamma \Bigl(\frac{n-3}{2}\Bigr)
\zeta_R (n-3) \biggr] \, . \label{free}
\end{eqnarray}
Here $a_k\equiv a_k(I,L)$, $\zeta_R(s) = \sum_{n=1}^{+\infty}
n^{-s}$ is a Riemann zeta function, $\zeta_L(s)=\sum_{\lambda}
\lambda^{-s}$ is a zeta function of an operator $L$, $\gamma$ is
the Euler constant. The first two terms on the r.h.s. of
(\ref{free}) follow from the $n=0$ term.

The term
\begin{equation}
-\frac{(4 \pi)^{3/2} \hbar a_0}{\beta^4} \frac{\pi^2}{90} =
 - V  \: \frac{{\rm tr} I}{ \hbar^3} \frac{\pi^2}{90}  T^4
\end{equation}
is the leading high temperature contribution to the free energy.

The classical limit terms due to the equality $\zeta_L(0) = a_3$
can be rewritten as follows:
\begin{equation}
T \Bigl( -\frac{1}{2} \zeta_L^\prime(0) + \zeta_L(0) \ln
\frac{\hbar}{2\pi T} \Bigr) = T \, \sum_\lambda \ln \frac{\hbar
\sqrt{\lambda}}{2\pi T} \, . \label{class}
\end{equation}
The terms on the l.h.s. of (\ref{class}) yield a renormalized
value of the terms on the r.h.s. of (\ref{class}), since the sum
on the righthandsight is generally divergent when the number of
modes is infinite.

The term with $a_4$ determines the part of the free energy that
appears due to one-loop logarithmic divergences and thus it
depends on the dimensional parameter $\mu$ as in the zero
temperature case.


\subsection{Chiral anomaly in four dimensions\\
 for MIT bag boundary conditions}

Consider the Dirac operator on an $n$-dimensional Riemannian
manifold
\begin{equation}
  \Dir =\gamma^\mu\left(\partial_\mu+ V_\mu+ iA_\mu\gamma^5- \frac{1}{8} [ \gamma_\rho,
  \gamma_\sigma ] \sigma^{[\rho \sigma]}_\mu \right)
  \label{A2}
\end{equation}
in external vector $V_\mu$ and axial vector $A_\mu$  fields. We
suppose that $V_\mu$ and $A_\mu$ are anti-hermitian matrices in
the space of some representation of the gauge group.
$\sigma^{[\rho \sigma]}_\mu$ is the spin-connection\footnote{The
spin-connection must be included even on a flat manifold if the
coordinates are not Cartesian.}.

The Dirac operator transforms covariantly under infinitesimal
local gauge transformations (the local gauge transformation is
$\Dir \to \exp (-\lambda )\Dir\exp (\lambda )$):
\begin{eqnarray}
&&\delta_\lambda A_\mu =[A_\mu ,\lambda ] \nonumber \\
&&\delta_\lambda V_\mu = \partial_\mu \lambda +[V_\mu ,\lambda ]
\nonumber \\
&&\Dir \to \Dir +[\Dir ,\lambda ] \label{gauge}
\end{eqnarray}
and under infinitesimal local chiral transformations (the local
chiral transformation is $\Dir \to \exp (i\varphi
\gamma_5)\Dir\exp (i\varphi \gamma_5 )$):
\begin{eqnarray}
&&\tilde\delta_\varphi A_\mu =\partial_\mu \varphi +[V_\mu
,\varphi ],
\nonumber \\
&&\tilde\delta_\varphi V_\mu =-[A_\mu ,\varphi ] ,\nonumber \\
&&\Dir \to \Dir + i\{ \Dir , \gamma^5 \varphi \} \,.
\label{chiral}
\end{eqnarray}
The parameters $\lambda$ and $\varphi$ are anti-hermitian
matrices.

We adopt the zeta-function regularization and write the one-loop
effective action for Dirac fermions at zero temperature as
\footnote{The one-loop effective action is proportional to Planck
constant $\hbar$, in what following we put $\hbar=1$.}
\begin{equation}
W=-\ln \det \Dir = -\frac 12 \ln \det \Dir^2 = \frac 12
\zeta_{\Dir^{2}}^\prime(0) +\frac 12 \ln (\mu^2) \zeta_{\Dir^{2}}
(0) \,,\label{detD}
\end{equation}
where
\begin{equation}
\zeta_{\Dir^{2}} (s)={\rm Tr} ( \Dir^{-2s} )\,, \label{zeta}
\end{equation}
prime denotes differentiation with respect to $s$, and ${\rm Tr}$
is the functional trace.

The following identity holds:
\begin{equation}
\zeta_{A}(s)= {\rm Tr} A^{-s} \Rightarrow \delta \zeta_{A}(s)=-s
{\rm Tr}((\delta A)A^{-s-1})\, . \label{zeta1}
\end{equation}

Due to the identity (\ref{zeta1})
\begin{equation}
\delta_\lambda \zeta_{{\Dir^{2}}}(s) =- \left( 2s {\rm Tr} (
[\Dir,\lambda ] \Dir^{-2s-1}) \right) = -2s \left( {\rm Tr} (
[\Dir^{-2s},\lambda ]) \right) =0\, ,
\end{equation}
so the effective action (\ref{detD}) is gauge invariant,
$\delta_\lambda W=0$.

The chiral anomaly is by definition equal to the variation of $W$
under an infinitesimal chiral transformation. Using (\ref{zeta1})
we obtain:
\begin{equation}
\tilde\delta_\varphi \zeta_{{\Dir^{2}}}(s) = -\left( 2is {\rm Tr}
( \{\Dir, \gamma^5\varphi\} \Dir^{-2s-1}) \right) = -4is \left(
{\rm Tr} (
 \gamma^5 \varphi \Dir^{-2s}) \right) \, ,
\end{equation}
and the anomaly reads
\begin{equation}
\mathcal{A}:=\tilde\delta_\varphi W = \frac 12
\tilde\delta_\varphi\zeta_{\Dir^{2}}^\prime (0)= -2 {\rm Tr} (i
\gamma^5 \varphi \Dir^{-2s} )\vert_{s=0} \,. \label{chian}
\end{equation}

The heat kernel is related to the zeta function by the Mellin
transformation:
\begin{equation}
{\rm Tr} (i \gamma^5 \varphi \Dir^{-2s} )=\Gamma (s)^{-1}
\int_0^\infty dt\, t^{s-1} K(i \gamma^5\varphi, \Dir^2, t) \,.
\label{zeta2}
\end{equation}
In particular, after the substitution of the heat kernel expansion
(\ref{hkexp}) into the formula (\ref{zeta2}) we obtain
\begin{equation}
\mathcal{A} =-2 a_n (i \gamma^5\varphi,\Dir^2) \,.\label{anomhk}
\end{equation}
The same expression for the anomaly follows also from the Fujikawa
approach \cite{Fujikawa:1979ay}.

We impose local boundary conditions:
\begin{equation}
\Pi_-\psi \BB =0,\qquad \Pi_-=\frac 12 \left( 1-\gamma^5 \gamma_n
\right) \,,\label{bagbc}
\end{equation}
which are nothing else than a Euclidean version of the MIT bag
boundary conditions \cite{bag}. For these boundary conditions
$\Pi_-^\dag =\Pi_-$, and the normal component of the fermion
current $\psi^\dag \gamma_n \psi$ vanishes on the boundary.

Since $\Dir$ is a first order differential operator it was enough
to fix the boundary conditions (\ref{bagbc}) on a half of the
components. To proceed with a second order operator $L=\Dir^2$ we
need boundary conditions on the remaining components as well. They
are defined by the consistency condition \cite{BG-Dirac}:
\begin{equation}
\Pi_-\Dir \psi \BB =0 \,,\label{conscond}
\end{equation}
which is equivalent to the Robin boundary condition
\begin{equation}
 \left(\nabla_n + S\right) \Pi_+ \psi \BB =0 \, ,\,\qquad \Pi_+=\frac 12 \left( 1+\gamma^5 \gamma_n
\right)
\end{equation}
 with
\begin{equation}
S= - \frac{1}{2}\Pi_+L_{aa} \,.\label{B11}
\end{equation}

In the paper \cite{lastpaper} the following expression for a
coefficient $a_4 (Q, L)$  with an hermitian matrix valued function
$Q$ and conditions (\ref{mixedbc}), $L_{ab}=0$ (flat boundaries),
$S=0$ was obtained:
\begin{eqnarray}
&&a_4(Q,L)=\frac 1{360} (4 \pi )^{
       -n/2} \Big\{ \iM \,{\rm tr}\,
       \big\{ Q(60{E_{ ; \mu}}^\mu+60 R E+180E^2 \nonumber \\
&&\qquad\qquad +30 \Omega_{\mu\nu}
      \Omega^{\mu\nu} +12 {R_{;\mu}}^\mu +
      5 R^2-2R_{\mu\nu}R^{\mu\nu}+2R_{\mu\nu\rho\sigma}R^{\mu\nu\rho\sigma})
\big\} \nonumber \\
&&\qquad\qquad+ \idM \,{\rm tr} \,
      \big\{ Q \{ 30 E_{;n} + 30 \chi E_{;n} \chi +
      90 \chi E_{;n} + 90 E_{;n}\chi  \nonumber\\
&&\qquad\qquad +18\chi\chi_{:a}\Omega_{an} + 12
\chi_{:a}\Omega_{an}\chi +
  18 \Omega_{an}\chi\chi_{:a}  -  12 \chi\Omega_{an}\chi_{:a}
  \nonumber\\ &&\qquad\qquad
         + 6 [\chi\Omega_{an}\chi, \chi_{:a}] + 54
[\chi_{:a}, \Omega_{an}] + 30 [\chi , \Omega_{an:a}] + 12 R_{ ;n}
+ 30 \chi R_{ ;n} \} +
 \nonumber\\ &&\qquad\qquad
       + Q_{ ;n}(- 30 E + 30 \chi E \chi + 90 \chi E
+ 90 E \chi - \nonumber \\
&&\qquad\qquad
 -18 \chi_{ :a} \chi_{ :a}   + 30 \chi_{:aa}
 - 6 \chi_{:a}\chi_{:a}\chi + 30
\chi R )+  30 \chi {Q_{ ;\mu}}^{\mu n}  \big\} \Big\} .
\label{Qa4bou}
\end{eqnarray}

To obtain the chiral anomaly in four dimensions\footnote{In two
dimensions ($n=2$) the boundary part of the chiral anomaly with
MIT bag boundary conditions is equal to zero \cite{lastpaper}.}
with MIT bag boundary conditions one has to calculate the
coefficient $a_4(Q, L)$ (\ref{Qa4bou}) with $L=\Dir^2$, $Q=
i\gamma^5 \phi$ and substitute it into (\ref{anomhk}). We define
$V_{\mu\nu} =
\partial_{\mu} V_{\nu} -
\partial_{\nu} V_{\mu} + [V_{\mu}, V_{\nu}]$,
$A_{\mu\nu} = D_{\mu} A_{\nu} - D_{\nu} A_{\mu}$, $D_{\mu} A_{\nu}
=
\partial_{\mu} A_{\nu} - \Gamma^{\rho}_{\mu\nu}A_{\rho}
+ [V_{\mu},A_{\nu}]$. The anomaly contains two contributions:
\begin{equation}
\mathcal{A}=\mathcal{A}_V + \mathcal{A}_b \,. \label{Ain4}
\end{equation}
In the volume part
\begin{eqnarray}
    \label{B16}
&&\mathcal{A}_V=
    \frac{-1}{180 \, (2 \pi)^{2} } \int_M \, d^{4}x \sqrt{g} \,
    {\rm tr} \varphi
     \Bigl( - 120 \, [D_{\mu}V^{\mu\nu},A_\nu] \nonumber\\
&&\qquad\qquad  +60 \, [D_{\mu}A_\nu,V^{\mu\nu}]
   - 60 \, D_{\mu}D^{\mu}D_{\nu}A^\nu
   + 120 \, \{\{D_{\mu}A_\nu , A^\nu\} , A^\mu\}\nonumber\\
&&\qquad\qquad   + 60 \, \{D_{\mu} A^\mu,A_\nu A^\nu \}
    + 120 \, A_\mu D_{\nu}A^\nu A^\mu
   + 30 \, [[A_\mu , A_\nu],A^{\mu\nu}]   \nonumber \\
&&\qquad\qquad    + \epsilon_{\mu\nu\rho\sigma}\
   \{ - 45 \, i\, V^{\mu\nu}V^{\rho\sigma}
    + 15 \, i \, A^{\mu\nu}A^{\rho\sigma}
   - 30 \, i \, (V^{\mu\nu}A^{\rho} A^{\sigma}
   + A^{\mu} A^{\nu} V^{\rho\sigma})
   \nonumber  \\
&&\qquad\qquad   - 120 \, i \, A^{\mu} V^{\nu\rho} A^{\sigma}
     + 60 \, i \, A^{\mu} A^{\nu} A^{\rho} A^{\sigma}\}
   - 60 \, (D_{\sigma}A_\nu) R^{\nu\sigma}
  + 30 \, (D_{\mu}A^\mu) R \nonumber \\
&&\qquad\qquad
   - \frac{15i}{8} \epsilon_{\mu\nu\rho\sigma}
\, R^{\mu\nu}{}_{\eta \theta}
    R^{\rho\sigma \eta \theta}
    \Bigr )
\end{eqnarray}
only the $DA - R$ terms seem to be new \cite{lastpaper} (for flat
space it can be found e.g. in \cite{Andrianov:1983fg}).

The boundary part
\begin{eqnarray}
&&\mathcal{A}_b=
    \frac{-1}{180 \, (2\pi)^2} \int_{\partial M} \, d^{3}x \sqrt{h}
    \,   {\rm tr} \Bigl(
    12 \, i \, \epsilon^{abc} \, \{A_b, \varphi \} D_a A_c
\nonumber \\
 &&\qquad\qquad   + 24 \{\varphi, A^a \} \{A_a, A_n\}
    -60 \, [A^a, \varphi] (V_{na} - [A_n, A_a]) \nonumber\\
&&\qquad\qquad  + 60 (D_n \varphi ) D_{\mu} A^{\mu}  \Bigr)
\label{banomaly}
   \end{eqnarray}
is new \cite{lastpaper}. It has been derived under the two
restrictions: $S=0$ and $L_{ab}=0$. Note, that in the present
context, the first condition ($S=0$) actually follows from the
second one ($L_{ab}=0$) due to (\ref{B11}).


\section*{Acknowledgements}
  V.M. thanks the organizers of the VIII Training Course
  in the Physics of Correlated Electron Systems and High-Tc
  Superconductors for partial support and hospitality in Vietri
  sul Mare (Salerno).
  V.M. is grateful to Dr. Adolfo Avella for his help and efforts
  that made the visit to Vietri possible.
  V.M. thanks Dmitri Vassilevich for his suggestions during the
  preparation of the paper. This work has been supported by the
  Ostpartnerschaften program of Leipzig University.



\bibliographystyle{aipprocl} 



\end{document}